# The analysis of the results of the Neutrino-4 experiment on search for sterile neutrino and comparison with results of other experiments

A.P. Serebrov, R.M. Samoilov,

*NRC "KI" Petersburg Nuclear Physics Institute, Gatchina, Russia*
E-mail: serebrov_ap@pnpi.nrcki.ru

Abstract

We present new results of measurements of reactor antineutrino flux and spectrum dependence on the distance in the range 6-12 meters from the center of the reactor core. Additional measurements were carried out and set of data to perform statistical analysis was almost doubled since the previous report. Using all collected data, we performed the model independent analysis on the oscillation parameters $\Delta m_{14}^2$ and $\sin^2 2\theta_{14}$. The method of coherent summation of results of measurements allows us to directly observe the effect of oscillations. We observed an oscillation effect in vicinity of $\Delta m_{14}^2 = (7.25 \pm 0.13_{stat} \pm 1.08_{syst})\text{eV}^2$ and $\sin^2 2\theta = 0.26 \pm 0.08_{stat} \pm 0.05_{syst}$. We provide a comparison of our results with results of other experiments on search for sterile neutrino. Combining the result of the Neutrino-4 experiment and the results of measurements of the gallium anomaly and reactor anomaly we obtained value $\sin^2 2\theta_{14} \approx 0.19 \pm 0.04$ (4.6σ). Also was performed comparison of Neutrino-4 experimental results with results of other reactor experiments NEOS, DANSS, STEREO, PROSPECT and accelerator experiments MiniBooNE, LSND and IceCube experiment.

Mass of sterile neutrino obtained from data collected in the Neutrino-4 experiment (in assumption $m_4^2 \approx \Delta m_{14}^2$) is $m_4 = 2.68 \pm 0.13$eV. Using the estimations of mixing angles obtained in other experiments and our new results we can calculate, within 3+1 neutrino model, masses of electron, muon, and tau neutrinos: $m_{\nu_e}^{eff} = (0.58 \pm 0.09)$eV, $m_{\nu_\mu}^{eff} = (0.42 \pm 0.24)$eV, $m_{\nu_\tau}^{eff} \leq 0.65$eV. Extended PMNS matrix for (3 + 1) model with one sterile neutrino is provided, neutrino flavor mixing scheme with sterile neutrino and global fit of reactor experiments.

## 1. Introduction

Experimental search for possible existence of neutrino oscillation into sterile state have been carried out for many years. That idea is under consideration in experiments carried out at accelerators, reactors, and artificial neutrino sources [1-23]. Sterile neutrino can be considered as a candidate for the dark matter.

The hypothesis of oscillation can be verified by direct measurement of the antineutrino flux and energy spectrum vs. distance at short 6 – 12m distances from the reactor core. We use method of relative measurements, which can be more precise. It requires a detector to be movable and spectrum sensitive. To detect oscillations to a sterile state, one needs to observe a deviation of flux-distance relation from $1/L^2$ dependence and alteration of the form of energy spectrum with distance. If such process does occur, it can be described at short distances by the equation:

$$P(\bar{\nu}_e \to \bar{\nu}_e) = 1 - \sin^2 2\theta_{14} \sin^2\left(1.27\frac{\Delta m_{14}^2[eV^2]L[m]}{E_{\bar{\nu}}[MeV]}\right) \quad (1)$$

where $E_{\bar{\nu}}$ is antineutrino energy in MeV, L – distance in meters, $\Delta m_{14}^2$ is difference between squared masses of electron and sterile neutrinos, $\theta_{14}$ is mixing angle of electron and sterile neutrinos. For the experiment to be conducted, one needs to carry out measurements of the antineutrino flux and spectrum as near as possible to a practically point-like antineutrino source.

## 2. Detector design

The detector scheme with active and passive shielding is shown in fig. 1. The liquid scintillator detector has volume of 1.8 m³ (5x10 sections having size of 0.225x0.225x0.85m³, filled with scintillator to the height of 70 cm). Scintillator with gadolinium concentration 0.1% was used in detector to register inverse beta decay (IBD) events $\bar{\nu}_e + p \to e^+ + n$. The method of antineutrino registration is to select a correlated pare of signals: prompt positron signal and delayed signal of neutron captured by gadolinium.

The neutrino detector active shielding consists of external and internal parts relative to passive shielding. The internal active shielding is located on the top of the detector and under it. The detector has a sectional structure. It consists of 50 sections – ten rows with 5 sections in each. The first and last detector rows were also used as an active shielding and at the same time as a passive shielding from the fast neutrons. Thus, the fiducial volume of the scintillator is 1.42 m³. For carrying out measurements, the detector has been moved to various positions at the distances divisible by section size. As a result, different sections can be placed at the same coordinates with respect to the reactor except for the edges at closest and farthest positions.

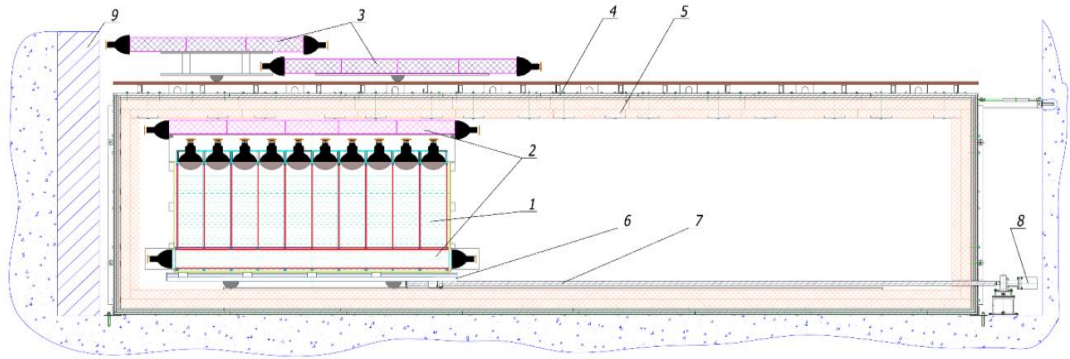

Fig. 1. General scheme of an experimental setup. 1 – detector of reactor antineutrino, 2 – internal active shielding, 3 – external active shielding (umbrella), 4 – steel and lead passive shielding, 5 – borated polyethylene passive shielding, 6 – moveable platform, 7 – feed screw, 8 – step motor, 9 – shielding against fast neutrons made of iron shot.

The measurements of fast neutrons and gamma fluxes in dependence on distance and reactor power were made before installing the detector into passive shielding [24]. Absence of noticeable dependence of the background on both distance and reactor power was observed. As a result, we consider that difference in signals (reactor ON - reactor OFF) appears mostly due to antineutrino flux from operating reactor. The signal generated by fast neutrons from reactor does not exceed 3% of the neutrino signal. The fast neutron background is formed by cosmic rays. The averaged over distance ratio of ON-OFF (antineutrino) signals to background is 0.5.

### 3. Measurements – the scheme of reactor operation and detector movements

The measurements with reactor under operation have started in June 2016 and were continued till June 2019, when reactor was stopped for renovation. From June 2019 till January 2020 the background has been measured. Measurements with the reactor ON were carried out for 720 days, and with the reactor OFF- for 417 days. In total, the reactor was switched on and off 87 times.

Measurements from September 2018 to July 2019 were carried out mainly in near positions to the reactor, where the signal to background ratio is significantly better. This measurement schedule made it possible to almost double (in comparison to the first stage of the experiment [24]) the amount of collected data in half the time and thus increase the statistical accuracy of measurements by factor 1.4. The scheme of reactor operation and detector movements is shown in fig. 2 at the top.

The ON-OFF difference is 223 events per day in distance range 6 – 9 m. Signal/background ratio is 0.54. To obtain antineutrino spectrum as difference ON-OFF background processes associated with cosmic radiation are subtracted. The measurements of fast neutrons and gamma fluxes in dependence on distance and reactor power were made before installing the detector into passive shielding. Absence of noticeable dependence of the background on both distance and reactor power was observed [24]. As a result, we consider that difference in signals (reactor ON - reactor OFF) appears mostly due to antineutrino flux from operating reactor. Thus, hereinafter ON-OFF count means antineutrino count.

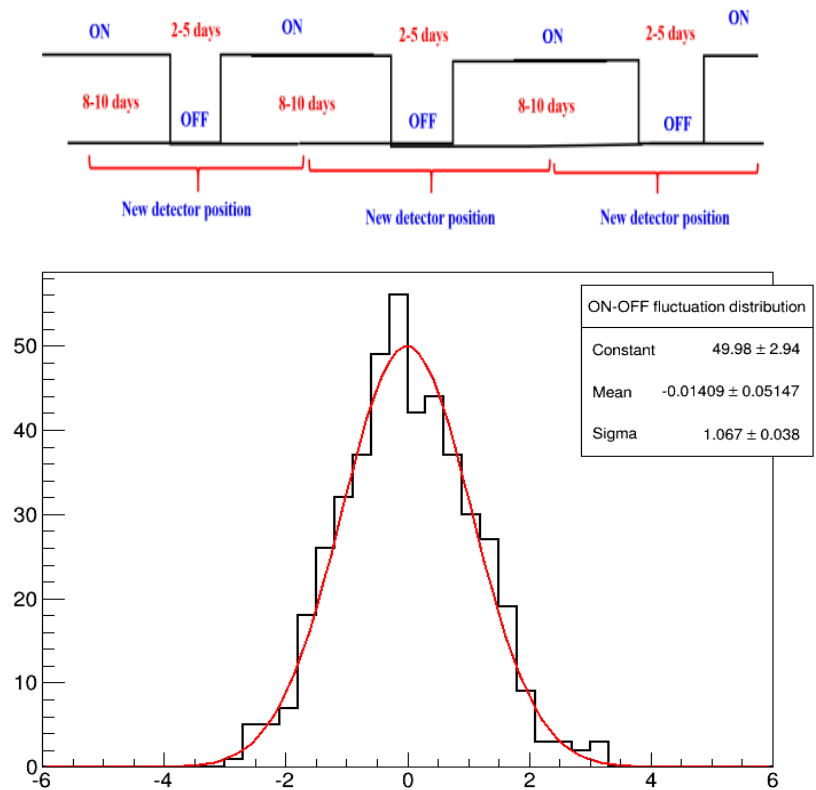

Fig. 2. Top - scheme of detector operation and detector movements; bottom - the distribution of deviations from average value of correlated events rates differences (ON-OFF) normalized on their statistical uncertainties.

The measurements of the background (OFF) and measurements with reactor in operation mode (ON) were carried out for each detector position within single measuring period. A reactor cycle is 8-10 days long. Reactor shutdowns are 2-5 days long and usually alternates (2-5-2-...). The reactor shutdowns in summer for a long period for scheduled preventive maintenance. The movement of the detector to the next measuring position takes place in the middle of reactor operational cycle. The stability of the results of measurements is characterized by distributions of ON-OFF difference fluctuations normalized on their statistical uncertainties, in measurements within one measuring period. The distribution is shown in fig.2 at the bottom. That distribution has the form of normal distribution, but its width exceeds unit by (7±4)%. This is a result of additional dispersion which appears due to fluctuations of cosmic background and impossibility of simultaneous measurements of the effect and background.

### 4. The matrix of measurements of the antineutrino flux dependence on distance and energy

The results of experimental measurements of the antineurino flux dependence on distance and energy of antineurino can be presented in the form of a matrix, which contains 216 elements, where $N_{ik}$ is difference of ON - OFF rates for i-th interval of energy and for k-th distance from reactor core. The energy spectrum is divided into 9 intervals of 500 keV, which corresponds to the energy resolution of the detector ±250 keV. The distnce step corresponds to the cell size of 23cm. In total there are 24 positions of antineurino flux measurements from 6.4m to 11.9m. Also more detailed data representations with devision into energy intervals 125keV and 250 keV were used.

### 5. Scheme of the experimental data analysis

There is a well-known problem of discrepancy between the experimental and calculated spectra, which also manifests itself in our experiment [23]. Therefore, method of the analysis of the experimental data should not rely on precise knowledge of the energy spectrum. Therefore, we propose model-independent method of data analysis, which employs equation (2), where the numerator is the rate of antineutrino events per $10^5$ s with a correction to geometric factor $L^2$ and denominator is the antineutrino events rate averaged over all distances:

$$(N_{ik} \pm \Delta N_{ik})L_k^2 / K^{-1} \sum_k^K (N_{ik} \pm \Delta N_{ik}) L_k^2 \ = \ \frac{\left(1-\sin^2 2\theta_{14} \sin^2\left(\frac{1.27\Delta m_{14}^2 L_k}{E_i}\right)\right)}{K^{-1}\sum_k^K \left(1-\sin^2 2\theta_{14} \sin^2\left(\frac{1.27\Delta m_{14}^2 L_k}{E_i}\right)\right)} \quad (2)$$

Equation (2) can be used to model-independent analysis of data because the left part includes only experimental data $k = 1, 2, ... K$ for all distances in the range 6.4-11.9 m, $K = 24$; $i = 1, 2, ... 9$ corresponding to 500 keV energy intervals in range 1.5 MeV to 6.0 MeV. The right part is the same ratio obtained within oscillation hypothesis. The left part is normalized to spectrum averaged over all distances; hence the oscillation effect is considerably averaged out in denominator if oscillations are frequent enough in considered distances range.

### 6. Monte Carlo calculation

In this section we present results of MC simulation in which we incorporated geometric configuration of the antineutrino source and detector including the sectioning.

The source of antineutrino with geometrical dimensions of the reactor core 42x42x35cm³ was simulated, as well as a detector of antineutrino taking into account its geometrical dimensions (50 sections of 22.5x22.5x85cm). The antineutrino spectrum of $U^{235}$ increased by function of oscillations $1 - \sin^2 2\theta_{14} \sin^2(1.27\Delta m_{14}^2 L_k/E_i)$ was used. Though it did not matter which particular energy spectrum of antineutrino we use since it is reduced in the equation (2). The most important parameter in this simulation was the energy resolution of the detector, which was ± 250 keV. Fig. 3 (left) shows the relationship of the oscillation pattern to the energy resolution of the detector. The oscillation curve corresponding to experimental energy resolution of the detector (± 250 keV) has to give the best fit of experimental data.

Fig. 3 (right) shows the simulated matrix of ratio $(N_{ik} \pm \Delta N_{ik})L_k^2/K^{-1}\sum(N_{ik} \pm \Delta N_{ik})L_k^2$ for calculations with $\Delta N_{ik}/N_{ik}$ equal to 1%, which is significantly better than the experimental value. One also can see a picture of the process of oscillations on the plane (E, L).

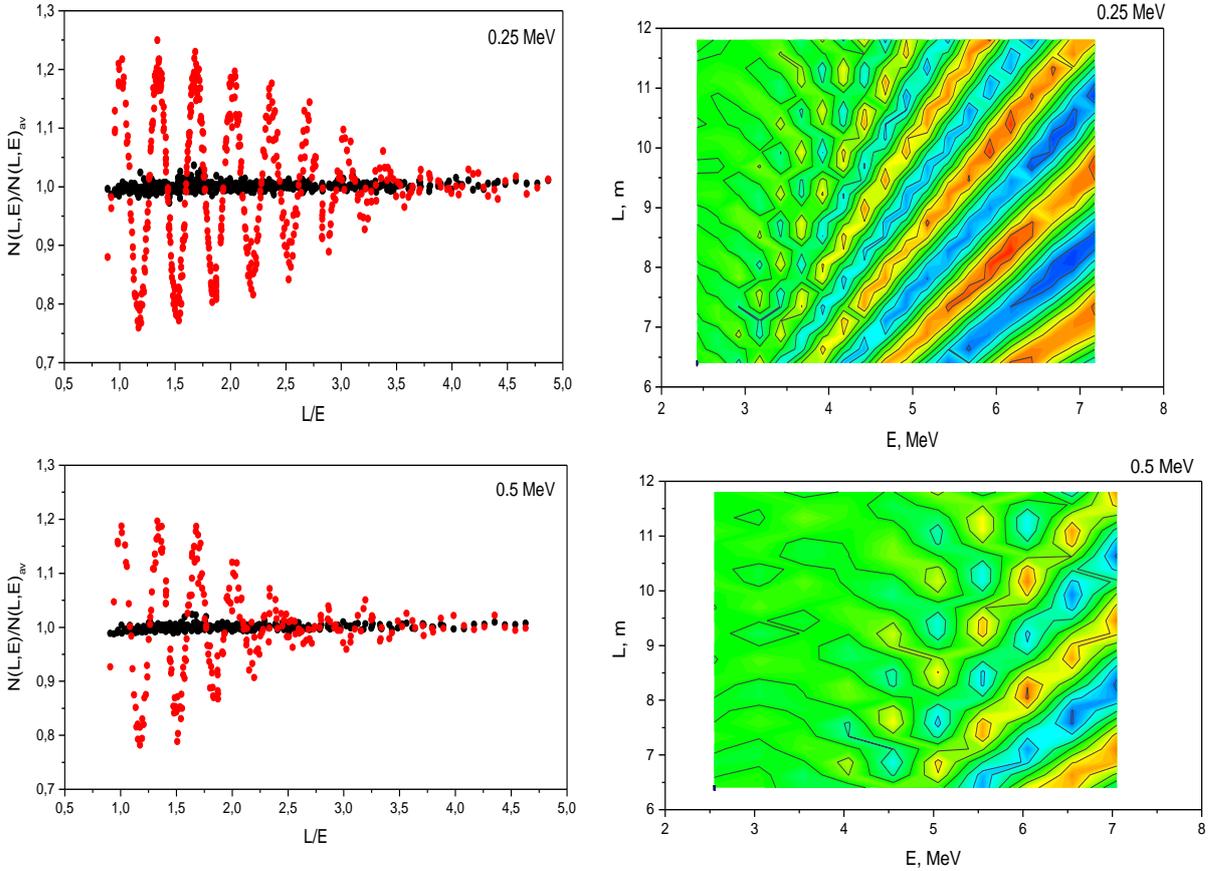

Fig.3. The simulated matrix of ratio $(N_{ik} \pm \Delta N_{ik})L_k^2/K^{-1}\sum(N_{ik} \pm \Delta N_{ik})L_k^2$ for various energy resolutions of detector.

The presented MC simulation reveals that resolution of the detector is extremely important for detecting the effect of the oscillations. Moreover, the oscillation effect could be extracted from data only by using the experimental dependence of ratio $(N_{ik} \pm \Delta N_{ik})L_k^2/K^{-1}\sum(N_{ik} \pm \Delta N_{ik})L_k^2$ on the parameter $L/E$. It should be noted that integration of the matrix over energy or distance significantly suppresses the ability to detect the effect of oscillations. Besides, the measurements in range 6 – 9 m are especially important, while measurements in range 9 – 12 m do not bring a significant contribution in sensitivity of the experiment, but they are used to correctly normalize the results of measurements.

### 7. Analysis of the experimental result search for oscillations

The matrix of measuremets incorporates data about the depedendence of antineutrino flux on distance and energy. The elements of the matrix $N_{i,k}$ represemts the difference ON-OFF signal in the i-th energy interval and k-th interval of distance to the center of the reactor core. This matrix should be compared with a calculated MC matrix, an example of such matrix is shown in fig. 3 on the right.

$$R_{ik}^{\text{exp}} = N(E_i, L_k)L_k^2 \bigg/ K^{-1}\sum_k^K N(E_i, L_k)L_k^2 = \frac{1 - \sin^2 2\theta_{14}\sin^2(1.27\Delta m_{14}^2 L_k/E_i)}{K^{-1}\sum_k^K(1 - \sin^2 2\theta_{14}\sin^2(1.27\Delta m_{14}^2 L_k/E_i))} = R_{ik}^{\text{th}}$$

If distance range for measurements is significantly larger than characteristic oscillation period denominator in $R_{ik}^{\text{th}}$ is simplified:

$$R_{ik}^{\text{th}} \approx \frac{1 - \sin^2 2\theta_{14} \sin^2(1.27\Delta m_{14}^2 L_k/E_i)}{1 - 1/2 \sin^2 2\theta_{14}} \xrightarrow[\theta_{14}=0]{} 1$$

Comparison of experimental matrix with calculated MC matrix can be done using $\Delta\chi^2$ method.

$$\sum_{i,k} \left(R_{ik}^{\text{exp}} - R_{ik}^{\text{th}}\right)^2 / \left(\Delta R_{ik}^{\text{exp}}\right)^2 = \chi^2(\sin^2 2\theta_{14}, \Delta m_{14}^2)$$

The results of the analysis of experimental data using $\Delta\chi^2$ method are shown in fig. 4-5.

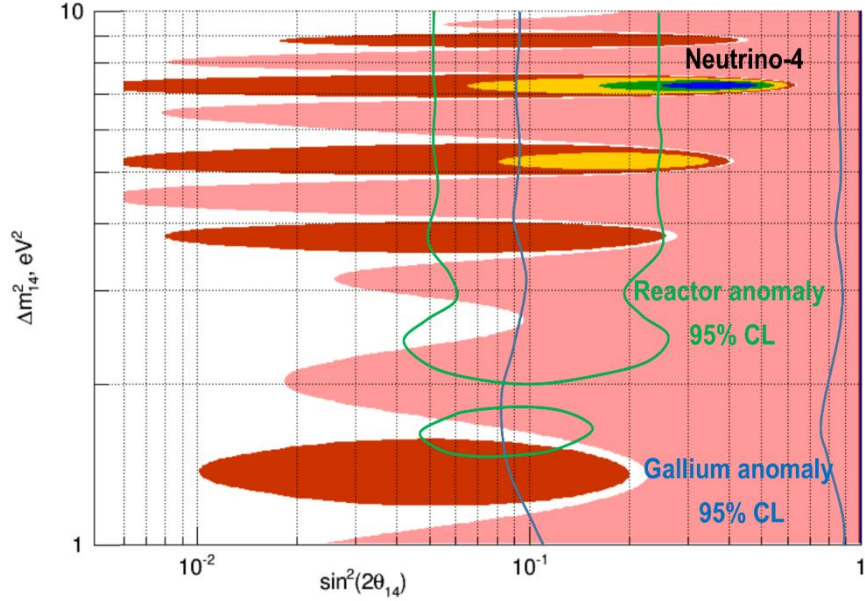

Fig. 4. Restrictions on parameters of oscillation into sterile state with 99.95% CL (pink), area of acceptable with 99.73% CL values of the parameters (yellow), area of acceptable with 95.45% CL values of the parameters (green), area of acceptable with 68.30% CL values of the parameters (blue).

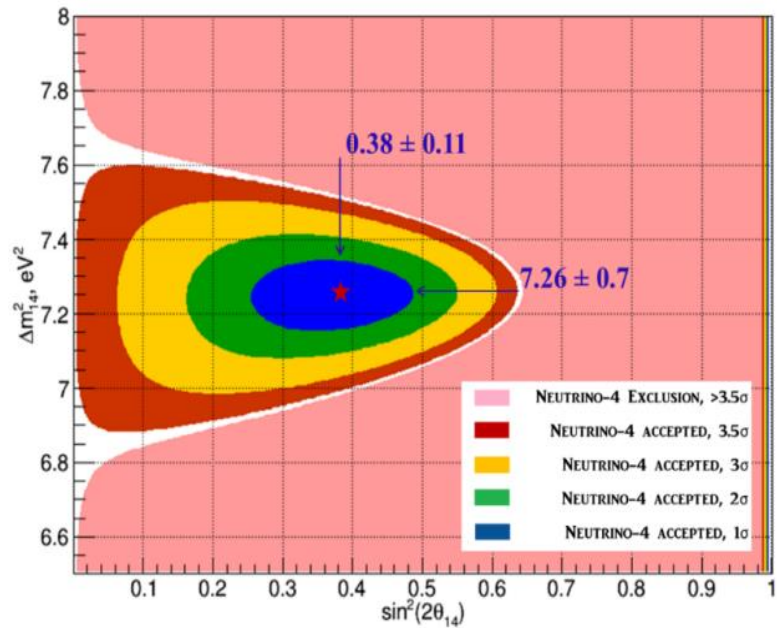

Fig. 5. The significantly magnified central area if prompt spectrum has 500 keV bin width.

The area of oscillation parameters colored in pink are excluded with CL more than 99.95% (>3.5σ). However, in area $\Delta m_{14}^2 = (7.26 \pm 0.7)\text{eV}^2$ and $\sin^2 2\theta_{14} = 0.38 \pm 0.11$ the oscillation effect is observed at 3.5σ CL, and this area is followed by a few satellites. Minimal value $\chi^2$ occurs at $\Delta m_{14}^2 \approx$ 7.26eV$^2$ for the case of data processing with energy spectrum divided on 500 keV bins. For the case of division on 125 keV, 250 keV, 500 keV bins with averaging these three data sampling oscillation effect is observing in area $\Delta m_{14}^2 = (7.25 \pm 0.13)\text{eV}^2$ and $\sin^2 2\theta_{14} = 0.26 \pm 0.08$ at 3.2σ CL.

The satellites appear due to effect of harmonic analysis where in presence of statistical noises along with base frequency we also can obtain frequencies equal to base frequency multiplied by integers and half-integers.

The stability of the results of the analysis can be tested. Using the obtained experimental data $(N_{i,k} \pm \Delta N_{i,k})$ one can perform a data simulation using randomization with a normal distribution around $N_{i,k}$ with dispersion $\Delta N_{i,k}$. Applying this method, 60 virtual experiments were simulated with results lying within current experimental accuracy. One can carry out the analysis described above for virtual experiments and average results over all virtual experiments. It was observed that obtained in this procedure exclusion area (pink area in fig. 4) coincide with one directly obtained from experimental data and oscillation parameters area is gathered around values $\Delta m_{14}^2 \approx 7.26\text{eV}^2$ and $\sin^2 2\theta_{14} = 0.38 \pm 0.11$

At last, one can simulate the experimental results with accuracy equal to experimental one, but in assumption of zero antineutrino oscillations. Obtained result reveals that amplitude of oscillations along horizontal axis, i.e. along the axis of parameter $\sin^2 2\theta_{14}$, is significantly reduced. It signifies that big amplitude of oscillations in fig. 4 indicates an existence of the oscillation effect. Sets of data simulated in assumption of zero oscillations and dispersion equal to accuracy of the experiment allow us to estimate sensitivity of the experiment at CL 95% and 99%. Obtained results can be used to compare sensitivity of our experiment with sensitivity of other experiments.

### 8. Analysis of the experimental result search for oscillations

As previously noted, the effect of oscillations can be revealed from the construction of the dependence of the experimental ratio $N_{ik}L_k^2/K^{-1}\sum N_{ik}L_k^2$ as function from L/E. Coherent sum of data with same L/E allows to demonstrate oscillation effect directly. Method $\Delta\chi^2$, using earlier for comparison E,L matrix with calculated one, allows to find the presence of oscillations and identifies optimal parameters. Using these optimal parameters, we construct an experimental ratio $N_{ik}L_k^2/K^{-1}\sum N_{ik}L_k^2$ as dependence from L/E and compare it with calculated dependence. Method $\Delta\chi^2$ is used again and optimality of parameters is checked.

More detailed analysis of the experimental data was performed with division of the energy spectrum using various intervals: 125 keV, 250 keV and 500 keV. This analysis was aimed to avoid fluctuations in the final result caused by usage of some particular system of data division. For this purpose, we used 24 distance points (with 23 cm interval) and 9 energy intervals (with 0.5MeV step) or 18 energy intervals (with 0.25MeV step) or 36 energy intervals (with 0.125MeV step). Corresponding matrices included 216, 432 and 864 elements. To form dependence of ratio $(N_{ik} \pm \Delta N_{ik})L_k^2/K^{-1}\sum(N_{ik} \pm \Delta N_{ik})L_k^2$ on parameter L/E we merged adjacent points into groups of 8, 16 and 32 correspondingly. At the next step the obtained L/E dependences were averaged and consequently the fluctuations of data divisions were averaged out.

The results of averaging of the data are shown in figure 6 (black squares). In purpose of comparison the results of analysis with interval 500 keV, which corresponds to energy resolution of the detector, are also presented (blue triangles). One can see that squares and triangles are statistically compatible.

A curve based on parameters $\Delta m_{14}^2 \approx 7.25\text{eV}^2, \sin^2 2\theta_{14} \approx 0.26$ provide a good fit of both sets of points. In analysis with energy interval 500 keV, which corresponds to energy resolution of the detector (blue triangles), the goodness of fit with such parameters is 45%, while fit with a constant equal to one (assumption of no oscillations) has the goodness of fit only 8%. We obtained $\chi^2/DOF = 17.1/17$ for the version with oscillations and $\chi^2/DOF = 30/19$ for the version without oscillations.

In analysis with averaging over data sets with energy intervals 125keV, 250 keV and 500keV (black squares) the fit with the given above parameters has the goodness of fit 28%, while fit with a constant equal to one (assumption of no oscillations) has the goodness of fit only 3%. We obtained $\chi^2/DOF = 20/17$ for the version with oscillation and $\chi^2/DOF = 32/19$ for the version without oscillation. To achieve the stability of the results we chose the analysis with averaging of the data. Corresponding confidence levels are shown in figure 7.

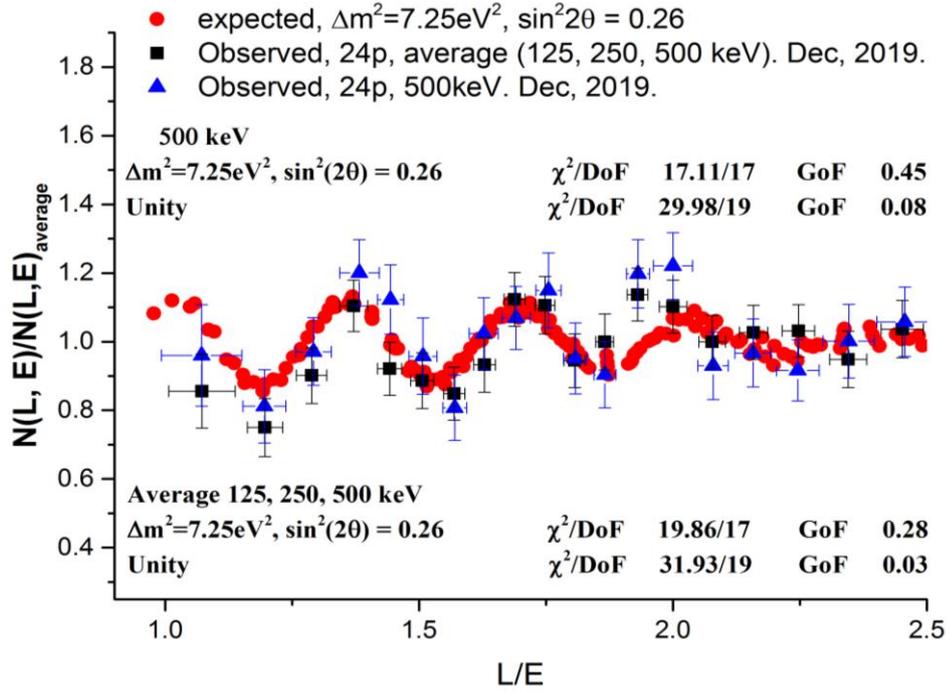

Fig. 6. The results of data analysis with energy interval 500 keV, which corresponds to energy resolution of the detector (blue triangles). The results of data analysis with averaging over energy intervals 125keV, 250keV and 500keV (black squares). Red dots – expected dependence at $\Delta m_{14}^2 = 7.25$ и $\sin^2 2\theta_{14} = 0.26$. Vertical errors are statistical, horizontal errors for blue triangles demonstrate the range of eight values for L/E ratio and average spread of L/E values for the averaged over 3 energy intervals (black squares) data sets. The period of oscillation for neutrino energy 4 MeV is 1.4 m.

Coherent summation is to combine data, obtained at different distances and different energy bins but close L/E ratio values. Combining data of 8 consecutive L/E ratio values we have some range in this ratio. Range forms horizontal errors demonstrated in figure 6 i.e. horizontal errors is difference between maximum and minimum in a row of 8 consecutive L/E ratio values.

Error $\Delta N_{ik}$ for $N_{ik}$ is statistical and is determined by the signal, correlated background and accidental background. During measurements with the reactor ON (measurement of the effect) and with reactor OFF (measurement of the correlated background) in the time window for correlated events search an interval from 100 to 300 μs is used to measure the accidental background, which is subtracted. The error for the obtained difference is RMS of statistical errors of the correlated count rate (from 0 to 100 μs) and the accidental background count rate.

The total error of the $\Delta N_{ik}$ neutrino events count rate is the RMS of the correlated events count rate errors when the reactor is ON (signal + correlated background) and OFF (correlated background only), which were mentioned earlier.

The $N_{ik}$ uses all the data collected at the k-th distance in the i-th energy bin for the entire measurement time. At the same time, as noted in section 3, an important role is played by the measurement scheme and the fact that the reactor SM-3 has short cycles (8-10 days) and frequent 0 MWt power time (reactor OFF) for 2-5 days. The distribution of the ON-OFF count rate in the entire energy range, normalized to its standard deviation shows width exceeds of only $(7 \pm 4)\%$, which is taken into account in subsequent data processing.

An error for $R_{ik}$ ratio is calculated as the sum of the relative errors of the numerator and denominator, and the numerator makes a major contribution, since the denominator is an average over all 24 points and has significantly higher accuracy.

For reasons of reliability of the final result, we choose the case of data processing with averaging. In this case oscillation effect is observed in vicinity of parameters $\Delta m^2_{14} = (7.25 \pm 0.13)eV^2$ and $\sin^2 2\theta_{14} = 0.26 \pm 0.08$ with confidence level (3.2σ).

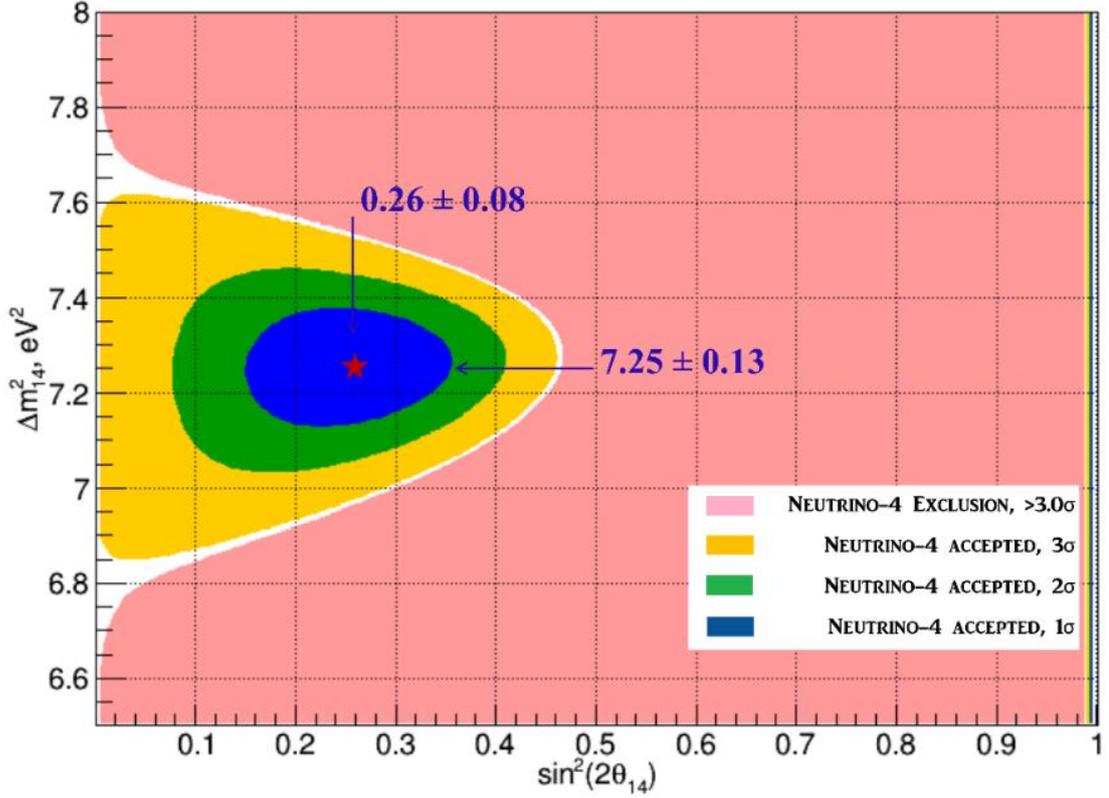

Fig. 7. Confidence levels of the area around oscillation parameters obtained as the best fit in case of averaging over three data sets.

## 9. Additional analysis of result validity.

It is often discussed that stricter limitations on the confidence level of the result can be obtained using the Feldman-Cousins method. In compliance Wilks theorem $\Delta\chi^2$ method is possible to apply successfully if effect is observed at the level of reliability 3σ more. The result of processing without taking into account systematic errors with an energy interval of 500 keV is $\sin^2 2\theta_{14} = 0.38 \pm 0.11(3.5\sigma)$, and when averaging data over 125keV, 250keV and 500keV is $\sin^2 2\theta_{14} \approx 0.26 \pm 0.08(3.2\sigma)$. Since the reliability of the effect we observe exceeds $3\sigma$, we do not consider it mandatory to use the Feldman-Cousins method and propose to do another additional analysis of our data.

Initial distribution of the count rate ON- OFF in the entire energy range was shown in figure 2 (bottom) and in fig. 8 (top) for obviousness. It shows a normal distribution determined practically by statistics. In fig. 8, we compare it with the distribution obtained for the ratio $R^{\text{exp}}_{ik}$ for the same dataset. It, as well as the distribution ON-OFF, normalized by $\sigma$. Figure 8 (bottom) shows the distribution of all 216 points over the L/E range from 0.9 to 4.7. You can see that the distribution $R^{\text{exp}}_{ik}$ already differs from normal ($\sigma = 1$, $\mu = 0$ and it normalized as $R^{\text{exp}}_{ik}$) due to the effect of oscillations. Value of the $\chi^2$/dof parameter is 25.9/16 which disfavors this function because of confidence level for this result is only 5%. It can be seen that this analysis uses initial data before processing for oscillation parameters Summing up, we would like to note that the effect of oscillations is manifested using three processing methods.
1. $\Delta\chi^2$ method at plane $(\sin^2 2\theta_{14}, \Delta m^2_{14})$,

2. Coherent summation method by variable L/E,
3. Analysis of distribution $R_{ik}^{exp}$ as opposed to normal distribution due to the effect of oscillations.

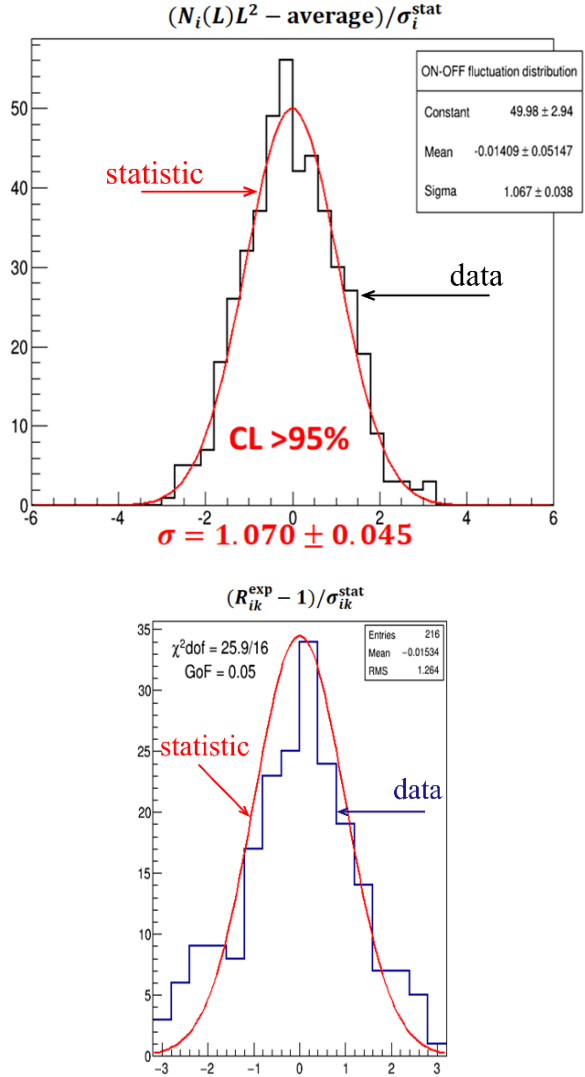

Fig. 8. Top - distribution of the count rate ON-OFF in the entire energy range, normalized by $\sigma$. Bottom - distribution $R_{ik}^{exp}$ of all 216 points over the L/E range from 0.9 to 4.7, normalized by $\sigma$

## 10. Systematic errors of the experiment

One of possible systematic errors of oscillation parameter $\Delta m_{14}^2$ is determined by accuracy of energy calibration of the detector, which is estimated to be ±250 keV. The relative accuracy of ratio L/E is determined by the relative accuracy of measurements of energy, because the relative accuracy of measurements of distance is significantly better. the relative accuracy of measurements of energy in the most statistically significant area of the measured neutrino spectrum 3-4 MeV is ±8%. Hence, possible systematic error of parameter $\Delta m_{14}^2$ is 0.6 eV$^2$, $\delta(\Delta m^2)_{syst1} \approx 0.6 eV^2$. Another systematic error of parameter $\Delta m_{14}^2$ can occur in data analysis performed with $\chi^2$ method because of additional regions around the optimal value $\Delta m_{14}^2 \approx 7.25 eV^2$. In particular, the closest regions have values 5.6eV$^2$ and 8.8eV$^2$, as can be seen from the fig. 9. However, its relative contribution to probability of occurring of this value is less than 9%. Hence, the possible systematic error can be estimated. As a result, the total systematic error of $\Delta m_{14}^2$ is $\delta(\Delta m^2)_{syst2} \approx 0.9 eV^2$. Finally, the obtained value of difference between masses of electron and sterile neutrino is:

$\Delta m_{14}^2 = 7.25 \pm 0.13_{st} \pm 1.08_{syst} = 7.25 \pm 1.09$.

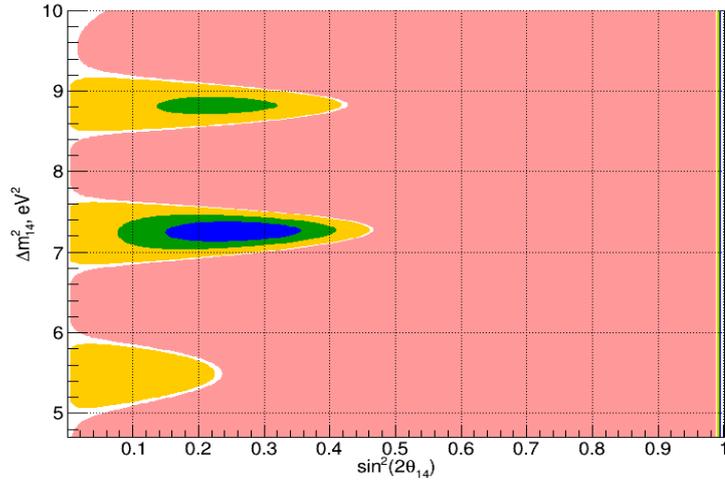

Fig. 9. Confidence levels of the additional regions around the area of the optimal oscillation parameters obtained as the best fit in case of averaging over three data sets

The systematic error of parameter $\sin^2 2\theta_{14}$ can occur in calculation of optimal value of $\sin^2 2\theta_{14}$ using $\chi^2$ method. The previously discussed analysis revealed that such error is possible. It was eliminated by more detailed analysis in which we used several energy intervals. That analysis with various energy intervals was amplified. As a result, it revealed that the standard deviation is less than 0.05, and that value should be considered as additional systematic error of the parameter $\sin^2 2\theta_{14}$. Therefore, $\delta(\sin^2 2\theta_{14})_{syst} \approx 0.05$ and mixing parameter is:

$$\sin^2 2\theta = 0.26 \pm 0.08_{stat} \pm 0.05_{syst}$$

Confidence level of statistical accuracy is $3.2\sigma$ and squared statistical and systematic errors is $\sin^2 2\theta = 0.26 \pm 0.09$.

**11. The dependence of the reactor antineutrino flux on distance in range 6-12 meters.**

Results of measurements of the difference in counting rates of neutrino events (reactor ON-OFF) are shown in fig. 10,11 as dependence of antineutrino flux on the distance to the reactor core. With this normalization, it is necessary to take into account the fact that already at a distance of 6 meters from the reactor core there is an averaging of the effect of oscillations for the energy spectrum integral. This leads to the well-known effect of neutrino flux deficiency, which is $1 - 1/2\sin^2 2\theta_{14}$ or 0.87 for $\sin^2 2\theta_{14} = 0.26$.

Thus, without making absolute measurements of the antineutrino flow from the reactor, we know the size of the deficit over long distances, taking the hypothesis of oscillations.
Fit of an experimental dependence with the law A/L² yields satisfactory result. Goodness of that fit is 22%. Corrections for finite size of reactor core and detector sections are negligible – 0.3%, and correction for difference between detector movement axis and direction to center of reactor core is also negligible – about 0.6%.

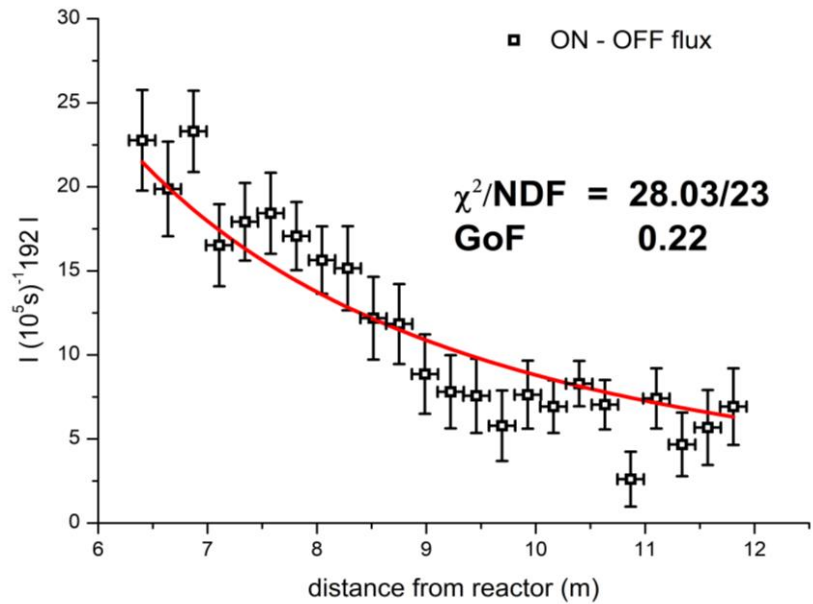

Fig. 10. Dependence of antineutrino flux on the distance to the reactor core – direct measurements.

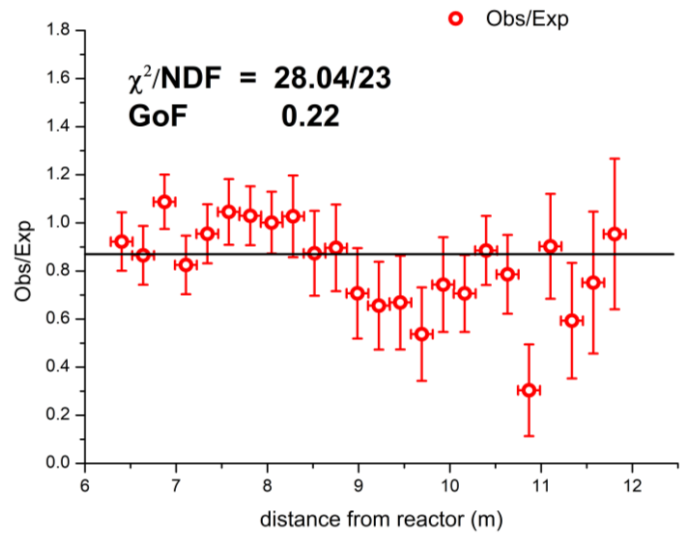

Fig. 11. Representation of experimental results in form of dependence of antineutrino flux on the distance to the reactor core normalized with the law A/L$^2$.

Figure 12 shows the dependence on the distance starting from the reactor core for the integral from the energy spectrum of the antineutrino flow calculated for $\Delta m_{14}^2 = 7.25$ и $\sin^2 2\theta_{14} = 0.26$. Four experimental points on this dependence correspond to intervals: 6-8 meters, 8-10 meters and 10-12 meters. Their position along the ordinate axis is $1 - 1/2\sin^2 2\theta_{14} = 0.87$, which reflects the deficit effect for the spectrum integral.

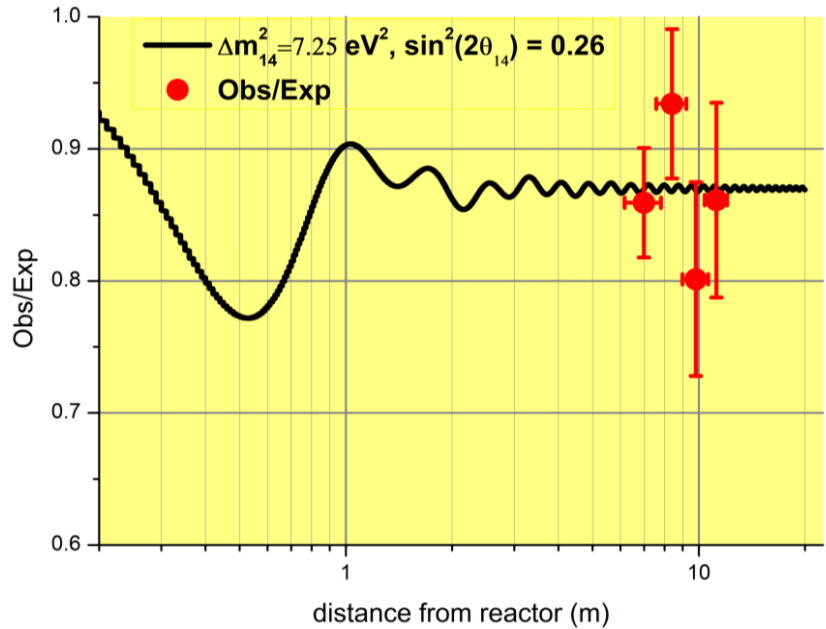

Fig. 12. Oscillation curve at the smallest distances calculated for $\Delta m_{14}^2 = 7.25$ and $\sin^2 2\theta_{14} = 0.26$ and measurement results in the range of 6-12 m.

## 12. Comparison of the result of experiment Neutrino-4 with reactor and gallium anomalies

In the Neutrino-4 experiment we measure the oscillation parameter $\sin^2 2\theta_{14}$, which is two times bigger than the deficiency of reactor antineutrino flux at large distance. In order to compare the results of Neutrino-4 experiment with results of measurements of reactor and gallium anomalies the obtained value of parameter $\sin^2 2\theta_{14}$ can be turned into the flux deficiency and vice versa. We will compare results in terms of oscillation parameter $\sin^2 2\theta_{14}$.

Figure 13 shows the famous oscillation curve of the reactor antineutrino with insertion of the picture of the oscillations obtained in the Neutrino-4 experiment with oscillation parameter $\sin^2 2\theta = 0.26 \pm 0.09\ (2.9\sigma)$. The neutrino deficiency called gallium anomaly (GA) [8,9] has oscillation parameter $\sin^2 2\theta_{14} \approx 0.32 \pm 0.10\ (3.2\sigma)$. The result of reactor antineutrino anomaly (RAA) [25-28] measurements is $\sin^2 2\theta_{14} \approx 0.13 \pm 0.05\ (2.6\sigma)$. Combination of these results gives an estimation for mixing angle $\sin^2 2\theta_{14} \approx 0.19 \pm 0.04\ (4.6\sigma)$.

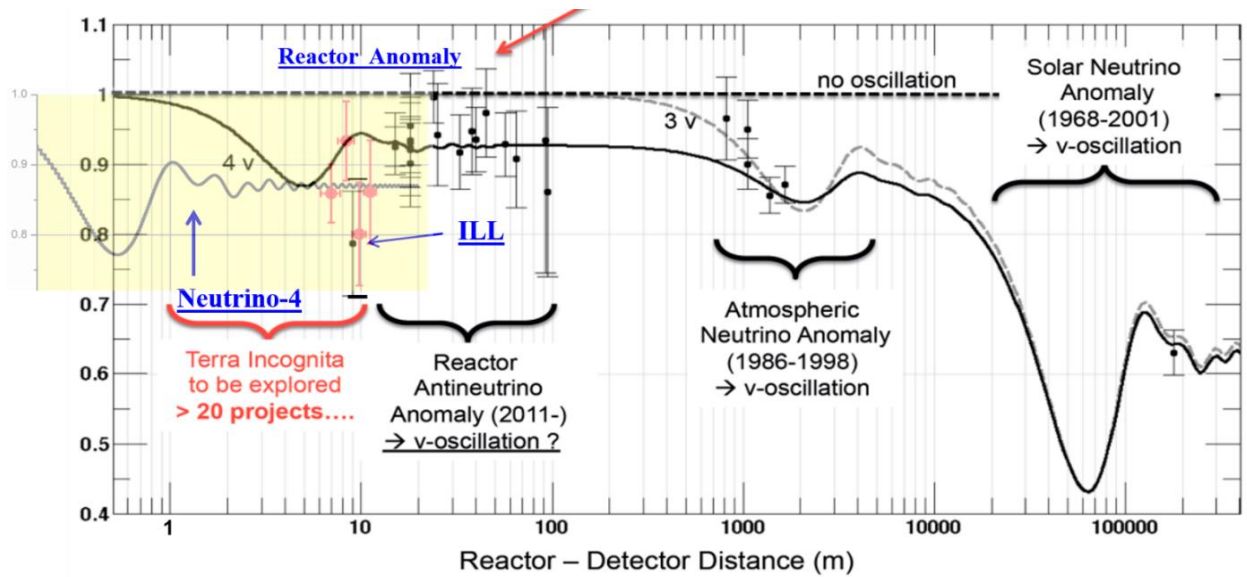

Fig.13. Reactor antineutrino anomaly [28] with oscillation curve obtained in experiment Neutrino-4.

**13. Comparison with other results of experiments at research reactors and nuclear power plants**

Figure 14 illustrates sensitivity of the experiment Neutrino-4 and other experiments DANSS [17], NEOS [18], PROSPECT [19] and STEREO [20]. In experiments on nuclear power plants sensitivity to identification of effect of oscillations with large $\Delta m_{14}^2$ is considerably suppressed because of the big sizes of an active zone. Experiment Neutrino-4 has some advantages in sensitivity to large values of $\Delta m_{14}^2$ owing to a compact reactor core, close minimal detector distance from the reactor and wide range of detector movements.

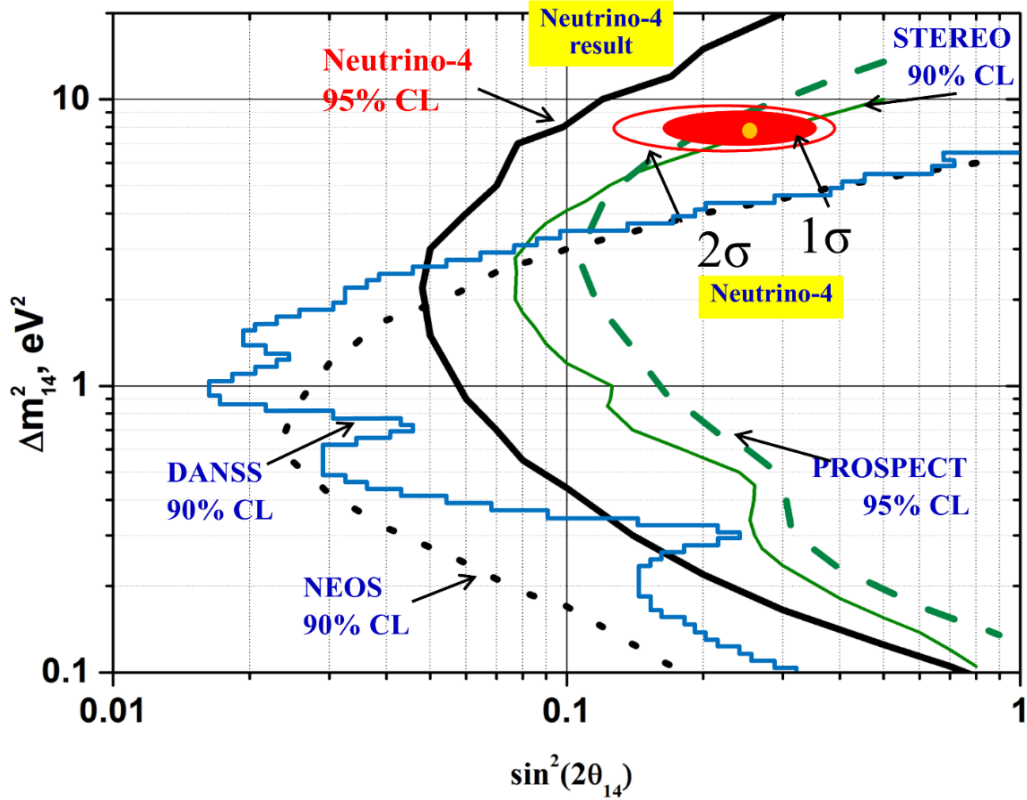

Fig. 14. Comparison of results of the Neutrino-4 experiment with results of other experiments – sensitivities of the experiments

Next highest sensitivity to large values of $\Delta m_{14}^2$ belongs to PROSPECT and STEREO experiment. Currently their sensitivities are two times lower than Neutrino-4 sensitivity, but it recently has started data collection, so possibly they will confirm our result after additional data taking. The experiment BEST started in August 2019 in BNO has good sensitivity at $\Delta m_{14}^2 > 5eV^2$ area [22].

It should be noted that without method of the coherent summation of data by L/E parameter, it is practically impossible to extract the effect of the oscillations. It should be noted that method of the coherent summation of data, by L/E parameter is necessary to demonstrate the real effect of oscillations. So far, the method of coherent summation of data by the parameter L/E at the short distance has been actively used only in experiment Neutrino-4. In fig. 15 it is shown comparison of planes of parameters (E, L) for experiments Neutrino-4, STEREO and PROSPECT. This may determine the difference in sensitivity between these experiments.

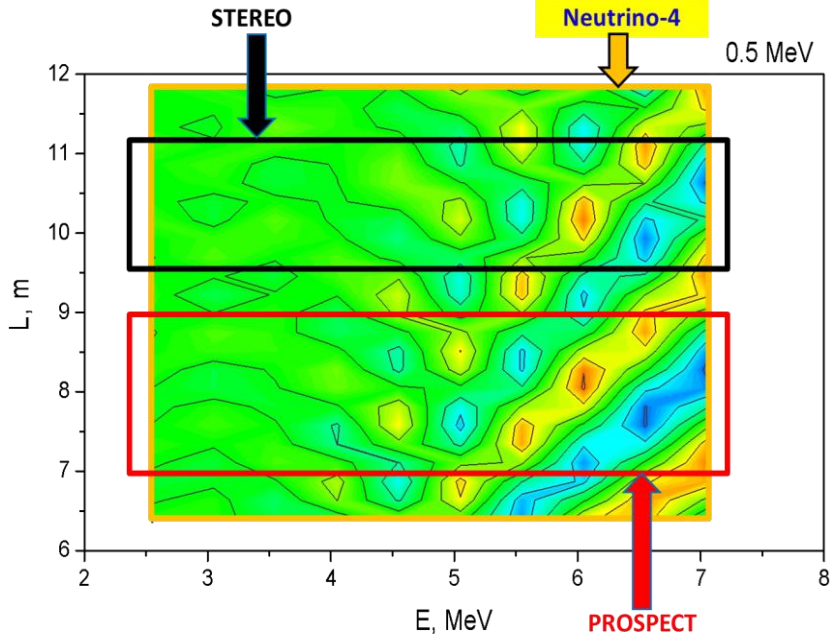

Fig. 15. Comparison of planes of parameters (E,L) in experiments Neutrino-4, STEREO and PROSPECT.

**14. The structure of 3 + 1 neutrino model and representation of probabilities of various oscillations**

In order to discuss a comparison with muon experiments we should start with structure of 3+1 neutrino model and representation of probabilities of various oscillations.

$$\begin{bmatrix} \nu_e \\ \nu_\mu \\ \nu_\tau \\ \nu_s \end{bmatrix} = \begin{bmatrix} U_{e1} & U_{e2} & U_{e3} & U_{e4} \\ U_{\mu 1} & U_{\mu 2} & U_{\mu 3} & U_{\mu 4} \\ U_{\tau 1} & U_{\tau 2} & U_{\tau 3} & U_{\tau 4} \\ U_{s1} & U_{s2} & U_{s3} & U_{s4} \end{bmatrix} \begin{bmatrix} \nu_1 \\ \nu_2 \\ \nu_3 \\ \nu_4 \end{bmatrix}$$

$$|U_{e4}|^2 = \sin^2(\theta_{14})$$
$$|U_{\mu 4}|^2 = \sin^2(\theta_{24}) \cdot \cos^2(\theta_{14})$$
$$|U_{\tau 4}|^2 = \sin^2(\theta_{34}) \cdot \cos^2(\theta_{24}) \cdot \cos^2(\theta_{14})$$

$$P_{\nu_e \nu_e} = 1 - 4|U_{e4}|^2(1-|U_{e4}|^2)\sin^2\left(\frac{\Delta m_{14}^2 L}{4 E_{\nu_e}}\right) = 1 - \sin^2 2\theta_{ee} \sin^2\left(\frac{\Delta m_{14}^2 L}{4 E_{\nu_e}}\right)$$

$$P_{\nu_\mu \nu_\mu} = 1 - 4|U_{\mu 4}|^2\left(1-|U_{\mu 4}|^2\right)\sin^2\left(\frac{\Delta m_{14}^2 L}{4 E_{\nu_\mu}}\right) = 1 - \sin^2 2\theta_{\mu\mu} \sin^2\left(\frac{\Delta m_{14}^2 L}{4 E_{\nu_\mu}}\right)$$

$$P_{\nu_\mu \nu_e} = 4|U_{e4}|^2|U_{\mu 4}|^2 \sin^2\left(\frac{\Delta m_{14}^2 L}{4 E_{\nu_e}}\right) = \sin^2 2\theta_{\mu e} \sin^2\left(\frac{\Delta m_{14}^2 L}{4 E_{\nu_e}}\right)$$

The relations of oscillations parameters required for comparative analysis of experimental results are:

$$\sin^2 2\theta_{ee} \equiv \sin^2 2\theta_{14}$$

$$\sin^2 2\theta_{\mu\mu} = 4 \sin^2 \theta_{24} \cos^2 \theta_{14} (1 - \sin^2 \theta_{24} \cos^2 \theta_{14}) \approx \sin^2 2\theta_{24}$$

$$\sin^2 2\theta_{\mu e} = 4 \sin^2 \theta_{14} \sin^2 \theta_{24} \cos^2 \theta_{14} \approx \frac{1}{4} \sin^2 2\theta_{14} \sin^2 2\theta_{24}$$

It is important that amplitudes of electron and muon oscillations with disappearance determines the amplitude $\sin^2 2\theta_{\mu e}$ in process with appearance of electron neutrinos in muon neutrino beam. It is an important relation which can be used for experimental verification of 3+1 neutrino model.

Experiments in which were obtained effects indicating process of oscillations in sterile state are Neutrino-4, reactor anomaly, gallium anomaly MiniBooNE, LSND, and IceCube.

## 15. Comparison of experiment Neutrino-4 results with results of the IceCube experiment

The comparison of results of the Neutrino-4 and the IceCube experiments is shown in fig. 16. In the IceCube experiment the best fit of data is obtained with parameters [29]:

$$\Delta m_{14}^2 = 4.47^{+3.53}_{-2.08} eV^2$$

$$\sin^2(2\theta_{24}) = 0.10^{+0.10}_{-0.07}$$

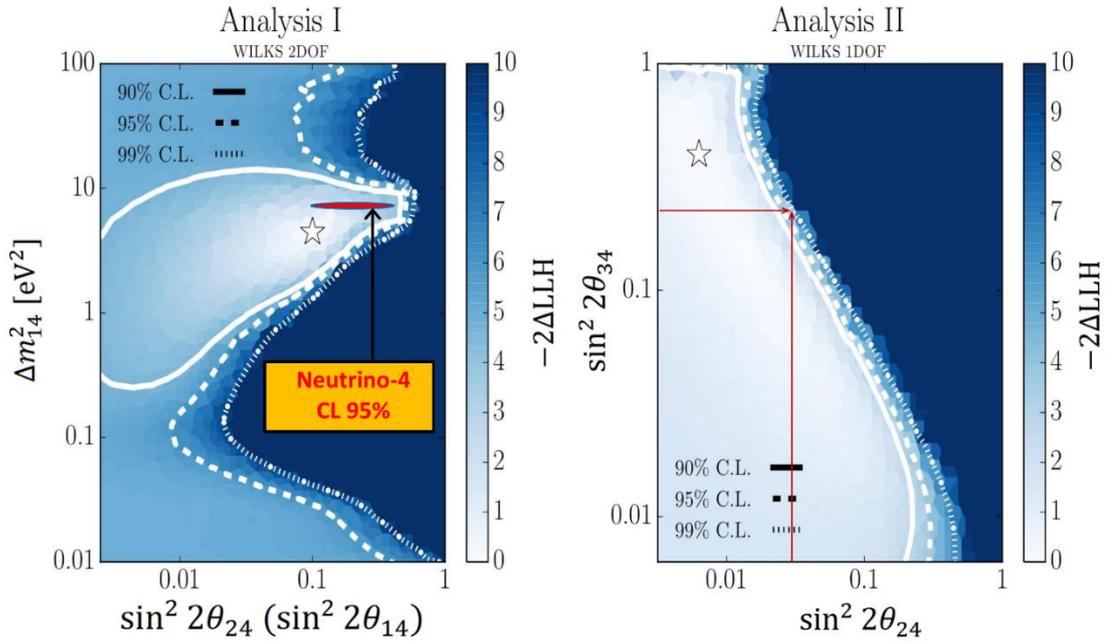

Fig.16. The comparison of Neutrino-4 and IceCube experimental results.

Values of parameter $\Delta m_{14}^2$ are in agreement within one standard deviation and values of $\sin^2 2\theta_{24}$ and $\sin^2 2\theta_{14}$ are in agreement within 1.3σ level, although the model 3 + 1 does not require this.

In [29] it is shown that lower limit of $\sin^2 2\theta_{24} \geq 0.03$ can be used to obtain upper limit of $\sin^2 2\theta_{34} \leq 0.21$ and that result can be used in order to estimate upper limit of tau neutrino mass.

## 16. Comparison of experiment Neutrino-4 results with results of accelerator experiments MiniBooNE and LSND

Furthermore, the interesting results can be obtained if we compare the results of the Neutrino-4 experiment with results of accelerator experiments (LSND[1] and MiniBooNE[2]). Using the data of that experiments represented in form of points on plots [30] we compared them with the results of Neutrino-4 experiment on the plane of parameters $\sin^2 2\theta_{\mu e}$ and $\Delta m_{14}^2$..

The experiments MiniBooNE and LSND are aimed to search for a second order process of sterile neutrino – the appearance of electron neutrino in the muon neutrino flux ($\nu_\mu \to \nu_e$) through an intermediate sterile neutrino. A comparison of $\sin^2 2\theta_{\mu e}$ obtained in MiniBooNE and LSND and $\sin^2 2\theta_{14}$ obtained in Neutrino-4 can be performed using results of the IceCube experiment: $\sin^2 2\theta_{24} \approx 0.03 \div 0.2$. Values of $\sin^2 2\theta_{\mu e}$ and $\sin^2 2\theta_{24}$, $\sin^2 2\theta_{14}$ are related by the expression: $\frac{1}{4}\sin^2 2\theta_{14}\sin^2 2\theta_{24}$.

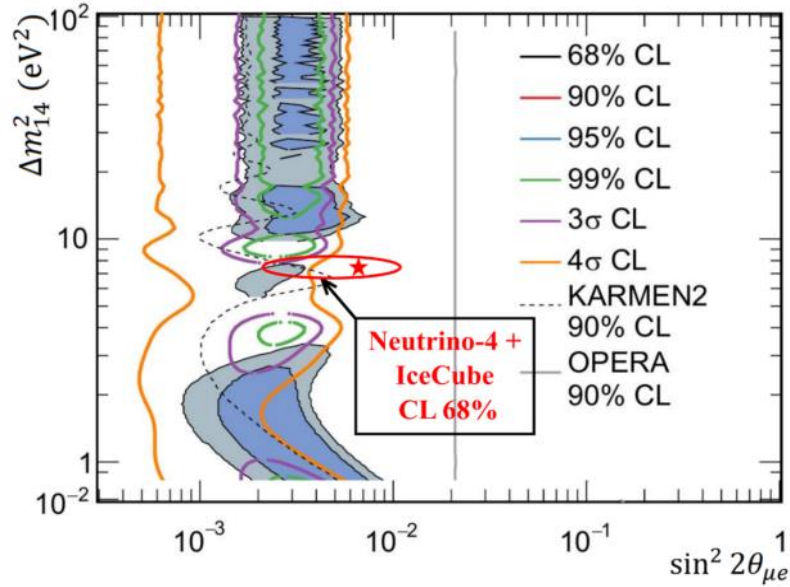

Fig. 17. Comparison of the results of the Neutrino-4 experiment with results of accelerator experiments MiniBooNE и LSND at the plane of parameters $\sin^2 2\theta_{\mu e}$ and $\Delta m_{14}^2$ and verification of the relation $\sin^2 2\theta_{\mu e} \approx \frac{1}{4}\sin^2 2\theta_{14}\sin^2 2\theta_{24}$

The calculated value of $\sin^2 2\theta_{\mu e}$ obtained after analysis of the Neutrino-4 and IceCube data is $\sin^2 2\theta_{\mu e} \approx 0.002 \div 0.013$ which is in agreement with value $\sin^2 2\theta_{\mu e} \approx 0.002 \div 0.006$ obtained in MiniBooNE and LSND. Therefore, mixing angles obtained with current experimental accuracy in experiments MiniBooNE, LSND, Neutrino-4, and IceCube are in agreement within 3+1 neutrino model (see fig. 17)

Comparison Neutrino-4 experiment results with results of MiniBooNE and LSND accelerator experiments on the plane of $\sin^2 2\theta_{\mu e}$, $\Delta m_{14}^2$ shown in figure 11 demonstrate there is a local minimum of $\Delta \chi^2$ distribution in area with large $\Delta m_{14}^2$ which corresponds to the area $\Delta m_{14}^2 \approx 7.25 \text{eV}^2$ of $\Delta \chi^2$ distribution of Neutrino-4 experiment.

## 17. Comparison with experiment KATRIN on measurement of neutrino mass.

The values of oscillation parameters obtained in the Neutrino-4 experiment can be used to estimate mass of the electron antineutrino, using general formulas for neutrino model [31,32] with extension to 3+1 model:

$$m_{\nu_e}^{eff} = \sqrt{\sum m_i^2 |U_{ei}|^2}$$

$$\sin^2 2\theta_{14} = 4|U_{14}|^2(1 - |U_{14}|^2)$$

$$|U_{14}^2| \ll 1;\ |U_{14}^2| \approx \frac{1}{4}\sin^2 2\theta_{14}$$

Limitations on the sum of mass of active neutrinos $\sum m_\nu = m_1 + m_2 + m_3$ from cosmology are in the range $0.54 \div 0.11 \text{eV}$ [33]. At the same time, knowing that $\Delta m_{14}^2 \approx 7.25 \text{eV}^2$, it is possible to consider that $m_4^2 \approx 7.25$ eV$^2$, and $m_1^2, m_2^2, m_3^2 \ll m_4^2$. Thus, the effective mass of the electron neutrino can be calculated by the formula: $m_{\nu_e}^{\text{eff}} \approx \sqrt{m_4^2 |U_{e4}|^2} \approx \frac{1}{2}\sqrt{m_4^2 \sin^2 2\theta_{14}}$.

With a more accurate consideration of this approximation using PMNS matrix data, the upper limit on the accuracy of the result does not exceed 10%.

It is necessary to make a little discussion here in connection with the known restrictions on the number of types of neutrinos and on the sum of the masses of active neutrinos from cosmology.

Depending on the scale of masses, sterile neutrinos can influence the evolution of the Universe and be responsible for the baryonic asymmetry of the Universe and the phenomenon of dark matter [34]. However, for sterile neutrinos with low mass and mixing angle, sterile neutrinos can be allowed to exist, which does not have a significant effect on cosmology [34]. Such sterile neutrinos practically do not thermalize in the primary plasma and leave it at an early stage.

Considering the above we can estimate sterile neutrino mass $m_4 = (2.68 \pm 0.13)$eV. In case of parameter $\sin^2 2\theta_{14} \approx 0.19 \pm 0.04 (4.6\sigma)$ obtained combining the results of the Neutrino-4 experiment and results of gallium anomaly measurements and more importantly using value $\Delta m_{14}^2 \approx (7.2 \pm 1.09)$eV$^2$ obtained for the first time in the Neutrino-4 experiment, we can make an estimation of the electron neutrino mass: $m_{\nu_e}^{\text{eff}} = (0.58 \pm 0.09)$eV. Obtained neutrino mass does not contradict the restriction on neutrino mass $m_{\nu_e}^{\text{eff}} \leq 1.1$ eV (CL 90%) obtained in the KATRIN experiment [35]. Moreover, the results of the determination of the sterile neutrino parameters make it possible to predict the value that can be obtained in the KATRIN experiment. Figure 18 shows sterile neutrino parameters constraints obtained in KATRIN experiment at the achieved accuracy and perspectives of its improvement [36].

In the same way we can use data about $\sin^2 2\theta_4$ obtained in the IceCube experiment to estimate muon neutrino mass: $m_{\nu_\mu}^{\text{eff}} = (0.42 \pm 0.24)$eV.

Finally, considering upper limit of $\sin^2 2\theta_{34} \leq 0.21$ we can calculate upper limit of tau neutrino mass $m_{\nu_\tau}^{\text{eff}} \leq 0.65$eV.

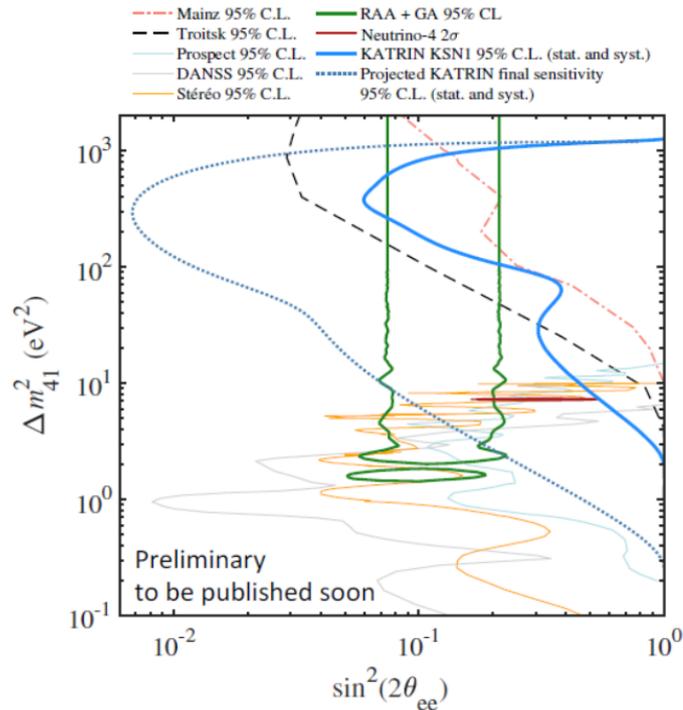

Fig. 18. Constraints on sterile neutrino oscillation parameters obtained from KATRIN and other experiments.

## 18. Comparison with neutrino mass constraints from experiments for neutrinoless double beta-decay search

In experiments for neutrinoless double beta decay, the Majorana neutrino mass is determined by the following expression:

$$m(0\nu\beta\beta) = \sum_{i=1}^{4} |U_{ei}|^2 m_i$$

This expression for the model 3 + 1 and with $m_1, m_2, m_3 \ll m_4$ assumption can be simplified: $m(0\nu\beta\beta) \approx m_4 U_{14}^2$. The numerical for this with Neutrino-4 and other experiments average result is shown below.

$$m(0\nu\beta\beta) = (0.13 \pm 0.03)\text{eV}$$

The best restrictions on the Majorana mass were obtained in the GERDA experiment [37]. In these experiments, the half-life of the isotope is measured, which depends on the Majorana mass as follows:

$$1/T_{1/2}^{0\nu} = g_A^4 G^{0\nu} |M^{0\nu}|^2 \frac{\langle m_{\beta\beta} \rangle^2}{m_e^2}$$

The upper limit for the half-life gives the upper limit for the Majorana mass:

Lower limit for $T_{1/2}^{0\nu} > 1.8 \times 10^{26}$ years (90% CL)

Upper limit for $m_{\beta\beta} < [80 - 182]$ meV

Further improvement of the accuracy of the double beta decay experiment may result in the detection of the Majorana mass or the closure of the Majorana neutrino. It should be noted that the results depend on the hierarchy of neutrino masses.

## 19. PMNS matrix for 3 + 1 model

The PMNS matrix for the (3 + 1) model with the sterile neutrino, whose parameters are determined in our Neutrino-4 experiment, in the experiments at the reactor and gallium anomaly, as well as in the experiment IceCube, is shown below:

$$U_{PMNS}^{(3+1)} = \begin{pmatrix} 0.824_{-0.008}^{+0.007} & 0.547_{-0.011}^{+0.011} & 0.147_{-0.003}^{+0.003} & 0.224_{-0.025}^{+0.025} \\ 0.409_{-0.060}^{+0.036} & 0.634_{-0.065}^{+0.022} & 0.657_{-0.014}^{+0.044} & 0.160_{-0.05}^{+0.08} \\ 0.392_{-0.048}^{+0.025} & 0.547_{-0.028}^{+0.056} & 0.740_{-0.048}^{+0.012} & < 0.229 \\ < 0.24 & < 0.30 & < 0.26 & > 0.93 \end{pmatrix}$$

Restrictions on $U_{si}$ values are obtained from matrix unitarity, provided that the sum of the squares of all four elements for each column does not exceed 1 more than one standard deviation. The scheme of mixing neutrino flavors with sterile neutrino for normal and inverted mass hierarchy is presented below (see fig. 19.).

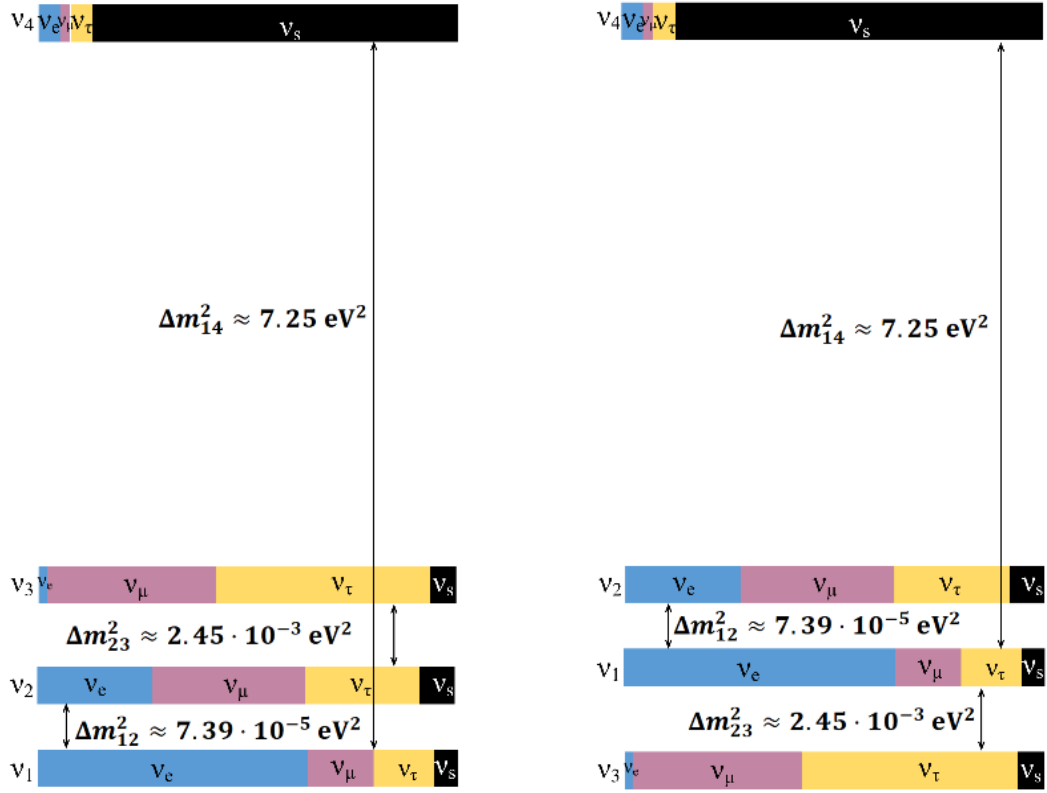

Fig. 19. Neutrino flavors mixing scheme including sterile neutrino for normal (on the left) and inverted mass hierarchy.

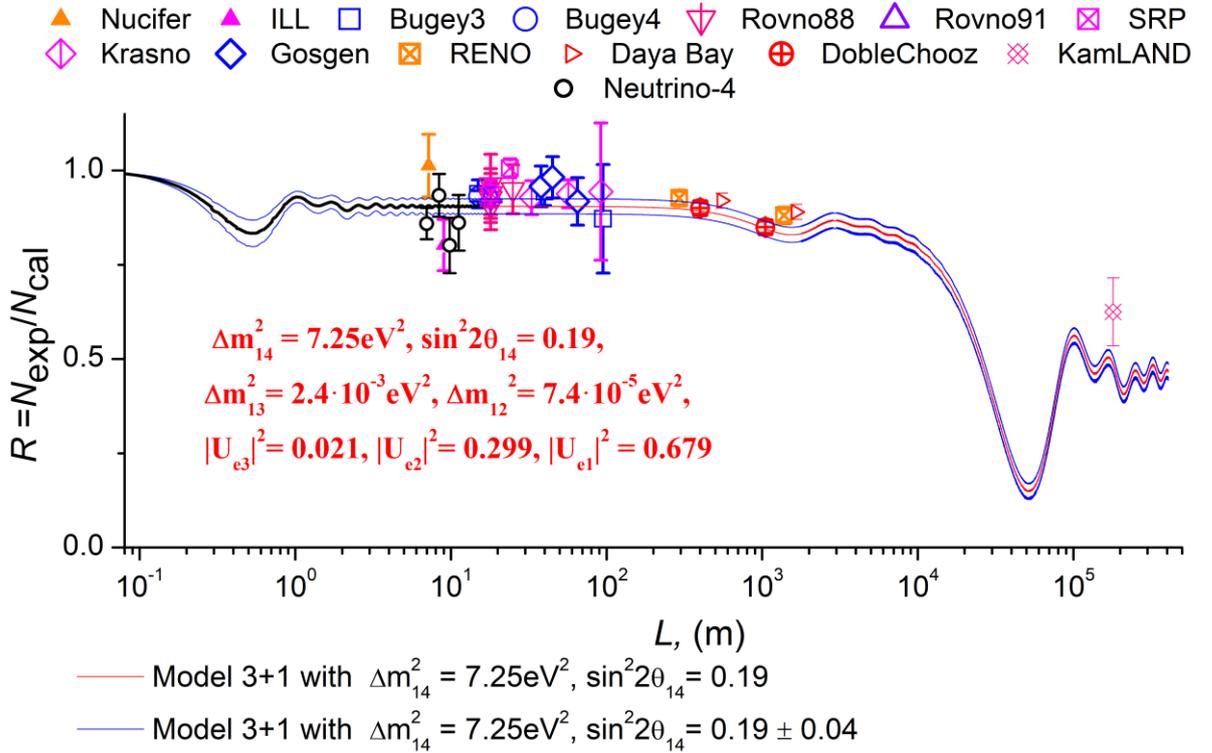

Fig. 20. Global picture of the reactor neutrino experiments supplemented with the effect of short baseline sterile neutrino oscillation.

At the end it is expedient to present global picture of reactor experiments supplemented with the effect of short baseline sterile neutrino oscillation (fig. 20).

## 20. Conclusions

The result of presented analysis of the Neutrino-4 experiment can be summarized in several conclusions.

1. Area of reactor and gallium anomalies with parameters $\Delta m_{14}^2 < 3 eV^2$ and $\sin^2 2\theta_{14} > 0.1$ is excluded at CL more than 99.7% (>3σ).

2. However, we observe an oscillation effect in vicinity of parameters. $\Delta m_{14}^2 \approx (7.25 \pm 1.09) eV^2$ and $\sin^2 2\theta = 0.26 \pm 0.08_{stat} \pm 0.05_{syst}$.

3. The obtained result can be compared with the results of other experiments aimed on search for sterile neutrino.

There are 5 types of experiments in which a deficiency in antineutrino (neutrino) registration is observed at $3\sigma$ CL

a) Neutrino-4 experiment,

b) In several reactor experiments, so-called reactor anomaly,

c) Experiments with neutrino source based on Cr51 (gallium anomaly).

d) Accelerator experiments MiniBooNE and LSND

e) the IceCube experiment

Table 1 presents results of various experiments: reactor anomaly, Neutrino-4 and gallium anomaly. Distribution of $\sin^2 2\theta_{14}$ parameter corresponding to these anomalies is shown in figure 20

**Table 1**

|  | Reactor anomaly | Neutrino-4 | Gallium anomaly |
|---|---|---|---|
| $\sin^2 2\theta_{14}$ | $0.13 \pm 0.05$ $(2.6\sigma)$ | $0.26 \pm 0.09$ $(2.9\sigma)$ | $0.32 \pm 0.10$ $(3.2\sigma)$ |
|  |  | $0.29 \pm 0.07$ $(4.3\sigma)$ ||
|  | $0.19 \pm 0.04$ $(4.6\sigma)$ |||

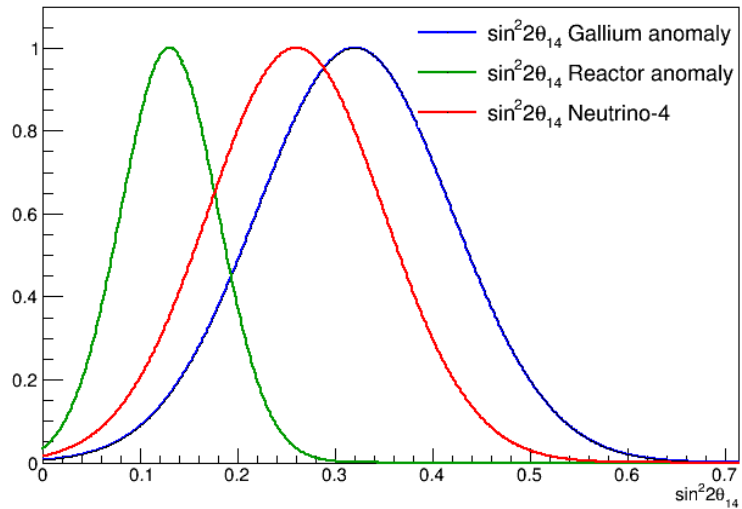

Fig. 20. Distribution of values $\sin^2 2\theta_{14}$ in GA, RAA, and Neutrino-4

4. Combining of these results gives estimation of mixing angle $\sin^2 2\theta_{14} \approx 0.19 \pm 0.04$ (4.6σ).

The correctness of Neutrino-4 result and RAA combining is questionable, but difference of these results is $0.13 \pm 0.09$ and it is only 1.4σ. Moreover RAA error not includes systematical error of reactor processes calculation, which is still under discussion.

5. Comparison of results obtained in the Neutrino-4 experiment with results of the IceCube experiment reveals a possible agreement of oscillation parameter from the Neutrino-4 experiment $\Delta m_{14}^2 \approx 7\text{eV}^2$ and oscillation parameter from the IceCube experiment $\Delta m_{14}^2 \approx 4.5\text{eV}^2$ within current accuracy of the IceCube experiment.

6. The comparison of results of the Neutrino-4 and IceCube experiments with accelerator experiments MiniBooNE and LSND at the plane of parameters $\sin^2 2\theta_{\mu e}$ and $\Delta m_{14}^2$ can be interpreted as agreement in oscillation parameter $\Delta m_{14}^2 \approx 7\text{eV}^2$. Calculated value of the $\sin^2 2\theta_{\mu e}$ from Neutrino−4 and IceCube experiments is $\sin^2 2\theta_{\mu e} \approx 0.002 \div 0.013$ and consistent with the value $\sin^2 2\theta_{\mu e} \approx 0.002 \div 0.006$ from MiniBooNE and LSND experiments.

7. Finally, from the analysis of Neutrino-4 result and results of other experiments discussed above one can make a conclusion about the possibility of existence of sterile neutrino with parameters $\Delta m_{14}^2 \approx (7.25 \pm 1.09)\text{eV}^2$ and $\sin^2 2\theta_{14} \approx 0.19 \pm 0.04 (4.6\sigma)$. Assuming that $m_4^2 \approx \Delta m_{14}^2$ we can estimate sterile neutrino mass $m_4 = (2.68 \pm 0.13)\text{eV}$.

8. The obtained values of oscillation parameters can be used to derive an estimation of the electron neutrino mass: $m_{\nu_e}^{\text{eff}} = (0.58 \pm 0.09)\text{eV}$.

9. Using the estimation of $\sin^2 2\theta_{24}$ obtained in the IceCube experiment and result $\Delta m_{14}^2 \approx (7.25 \pm 0.7)\text{eV}^2$ of the Neutrino-4 experiment we can estimate the muon neutrino mass to be $m_{\nu_\mu}^{\text{eff}} = (0.42 \pm 0.24)\text{eV}$ and upper limit of $\sin^2 2\theta_{34} \leq 0.21$ can be applied to estimate upper limit of tau neutrino mass: $m_{\nu_\tau}^{\text{eff}} \leq 0.65\text{eV}$

An illustration of estimations of masses of electron neutrino, muon neutrino, tau neutrino, and sterile neutrino are shown in fig. 21. The sterile neutrino determines masses of other neutrinos through mixing angles $\theta$ at level $0.1 \div 0.2$ and less.

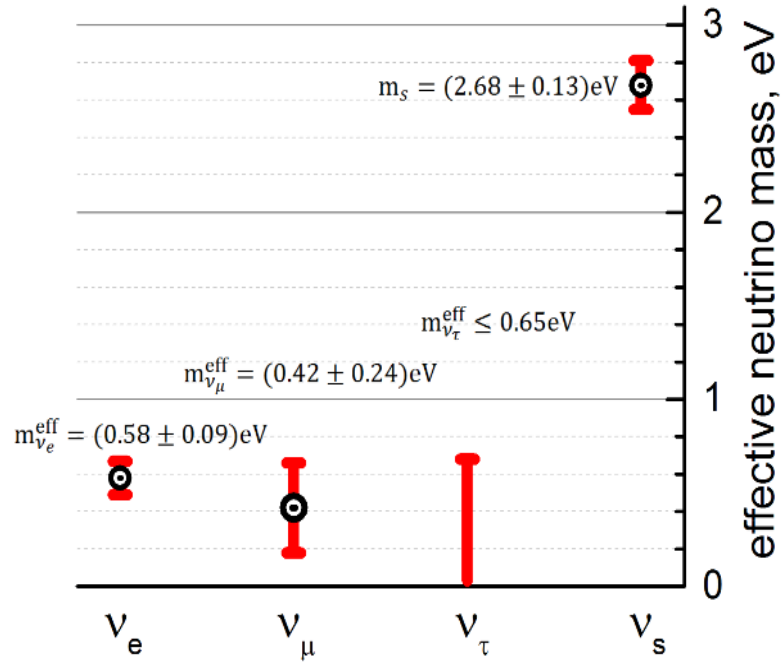

Fig. 21. The estimations of neutrino masses.

It should be noted that the sum of the effective masses of active neutrinos $m_{\nu_e}^{\text{eff}} + m_{\nu_\mu}^{\text{eff}} + m_{\nu_\tau}^{\text{eff}}$ is not directly related to cosmological estimates for the sum of masses $m_1 + m_2 + m_3$.

10. The PMNS matrix for four flavors together with sterile neutrino is presented. The parameters of matrix are determined in our Neutrino-4 experiment, in experiments on reactor and gallium anomaly, as well as in experiment IceCube.

The final confirmation of existence of sterile neutrino requires a result obtained with $5\,\sigma$ CL. We plan to create second neutrino laboratory at SM-3 reactor and new detector with three times higher sensitivity.

**Acknowledgements**

The authors are grateful to the Russian Science Foundation for support under Contract No. 20-12-00079. Authors are grateful to M.V. Danilov, V.B.Brudanin, V.G.Egorov, Y.Kamyshkov, V.A.Shegelsky, V.V. Sinev, D.S. Gorbunov and especially to Y.G.Kudenko for beneficial discussion. Also, authors would like to thank C. Rubbia for useful questions. The delivery of the scintillator from the laboratory headed by Prof. Jun Cao (Institute of High Energy Physics, Beijing, China) has made a considerable contribution to this research.